\documentstyle[preprint,aps,tighten,epsf]{revtex}
\begin{document}
\draft
\title{Width of phonon states  on defects  of various dimensions}
\author{L. A. Falkovsky \footnote{e-mail: falk@itp.ac.ru}}
\address{Landau Institute for Theoretical Physics, 117337 Moscow, Russia}
\maketitle

\begin{abstract}
We consider the localized phonon states created by   defects
of various geometries near the edge of an optical-phonon branch. 
The averaged Green's function 
 is calculated to study the Raman line shape. The phonon scattering 
 by the defects induces broadening and   line shape  asymmetry.
 The contribution of localized states to Raman spectra has a form of
 shoulder with a width proportional to the square root of 
 defect concentration. 
\end{abstract}


\section {Introduction}
The interaction of elementary excitations with
defects is an old problem of solid state physics. For the case of optical
phonons, it has attracted considerable attention in recent years
because  of the Raman scattering applications for determination
of the isotopic composition \cite{GRZ},  for
detection of   stacking faults in  polytypes \cite{NNH,RHL},
for investigations of residual strain
and strain relaxation at interfaces in semiconductors \cite{FBC}.
 
 A refinement of these effects takes the interactions  among phonons
 and  with defects into account. 
  
  (i) {\it The phonon decay} into
 two (or more) acoustic or optical phonons is possible even at zero
 temperature. These  unharmonic processes give rise to 
 the so-called natural width 
 $\Gamma^{\text{nat}}\sim \omega_0\sqrt{m/M}$, where $\omega_0$ is the
 optical-phonon frequency, $m$ and $M$ are the electron and atom masses.
 In spite of the small value $\sqrt{m/M}\sim 10^{-2}$, the decay processes
 give the observable  linewidth in Raman spectra and the quite small
 mean-free path $r_{\gamma}\sim a\sqrt{\omega_0/\Gamma^{\text{nat}}}\sim 10 a$,
 where $a$ is the lattice constant. The  decay cross section 
  depends weakly on the phonon frequency
 since no singularities exist in the final density of states for that processes. 
 Therefore, the natural optical-phonon width  
 $\Gamma^{\text{nat}}$ can be considered for given material 
 as constant up to room temperatures and
 for the  phonon frequency varying in a small interval of the order of
 $\Gamma^{\text{nat}}$. Then we obtain for perfect crystals the Lorentzian
 line shape of Raman spectra.
   
  (ii) {\it The phonon  interaction with defects} should be considered in two
 aspects. First, it gives rise to  elastic phonon scattering and makes
 an additional contribution to the phonon width $\Gamma$. This contribution
 (linear in defect concentration $c$)
 can be evaluated \cite{Tam} using the golden rule of quantum mechanics.
  The scattering by defects, being elastic, conserves  singularities of
 the phonon density of states. This is important for the Raman scattering, 
 because we are interested in phonons with small wave vectors. For
 the short-range potential of defects with the radius   of the order of 
 $a<r_{\gamma}$, the singularity  arises also in the width
 $\Gamma$ as  function of phonon frequency $\omega$ 
 at the phonon branch edge, i.e., at the maximum (or minimum) value $\omega_0$.   
 Thus we obtain the asymmetrical Raman line shape,  that was observed in 
 experiments with 3$C$-SiC \cite {FBC}. 
 This asymmetry is sensitive to the defect geometry,
 which may have the point, line or plain form \cite{Fan}.
 Secondly,
 the interaction with defects can result in appearance of new modes.
 For instance, the mode forbidden by  selection rules in  perfect crystals  
 at given experimental conditions 
 has been observed in  crystals with the stacking faults 
 which give an example of
 plain defects in SiC polytypes \cite{NNH}. Some additional modes were
 found also in  computer simulations for the sequence of number 
 layers with the stacking faults \cite{RHL}. 
 In addition, the localized mode can be associated with defects \cite{Li}.  
 
 The problem of  localized states on the single
 isotopic defect has the exact solution. However, the case of the low, but
 finite defect concentration involves  serious difficulties 
 \cite{GP,Bo}. That is
 because the solution should be self-consistent for the large distances  
 essential near the edge of phonon
 branches. If  the natural width is ignored, the answer  obtained
 in the series expansion of the phonon self-energy is the following. 
 The edge of phonon branches is displaced linearly in $c$
  and  the band of localized states arises with a width of the order of
  $c^2$. This $c^2$ behavior is explained by fluctuations in the spatial
  distribution of defects.

The aim of the present paper is to consider (taking the natural width
into account) the situation
when the  states localized on defects can exist.  
We are especially interested in the case when the localized states
appear near an edge of  optical-phonon
branches. Broadening of localized states is found just in the 
lowest order in the defect concentration $c$, if this additional width
is larger than the natural phonon width. This unusual behavior is
conditioned by the resonance interaction of localized states with the close
edge of the extended phonon branch.
The defects of different geometries are treated 
such as point  defects, line defects involved by dislocations,
 and plain defects  produced by stacking faults or 
 by the crystallite boundaries.
  They give the different    broadening and asymmetry
 of the Raman  line shape.  
The preliminary results of the work were published in Ref. \cite{Falk}.
 
\section {Interaction of phonon modes with defects}
We solve the equation for the phonon Green's function in the form:
\begin{equation} \label{gfe}
D({\bf r}, {\bf r}',\omega) = D_0({\bf r} - {\bf r}', \omega) 
- \int d{\bf r}'' D_0({\bf r} - {\bf r}'', \omega) U({\bf r}'')
D({\bf r}'', {\bf r}', \omega),
\end{equation}
where the interaction with defects located in points ${\bf r}_n$ is
\[U({\bf r})=\sum_{n}u({\bf r}-{\bf r}_n)
 = \sum_{{\bf n,q}}
u({\bf q}) e^{i{\bf q}({\bf r} - {\bf r}_n)}. \]

For an isotopic  defect, $u({\bf q})$ is constant: 
$u({\bf q})=u_0=(M_0 - M)a^3/M_0$, where $M$ and $M_0$ are
the masses of the defect atom and the host atom,
respectively, $a$ is the lattice constant. 
In the following, the large distances from the defect 
on the atomic scale are important. 
 Then for any type of defects, we can  set  the  Fourier component $u({\bf q})$ 
 equal to the constant $u_0$
  supposing  the short-range potential. For the same reason
we   expand  the phonon spectrum near
 the Brillouin-zone center taking
the Green's function in the absence of defects as
$$D_0(k,\omega)=
(\omega_0^2-s^2k^2-\omega^2-i\omega\Gamma^{\text{nat}})^{-1},$$
where the natural phonon width $\Gamma^{\text{nat}}$ is connected to decay into
 acoustic or optical phonons. 

The imaginary part of $D({\bf k},\omega)$
gives the differential density of squared
frequencies. 
In addition, the Raman intensity $I(\omega)$ for phonon excitation 
with frequency $\omega$ and momentum ${\bf k}$ is proportional to
$\text{Im}D({\bf k},\omega)$
which gives the Lorentzian line shape  in the absence of defects
$$ I(\omega)\sim \text{Im}D_0({\bf k},\omega)=\frac{\omega\Gamma^{\text{nat}}}
{(\omega_0^2-s^2k^2-\omega^2)^2+(\omega\Gamma^{\text{nat}})^2},$$
if $\Gamma^{\text{nat}}$ is assumed  to be small in comparison 
to the frequency $\omega_0$.

In the presence of defects, the Green's function must be averaged over defect
positions. It is convenient to use  the well-known diagram technique
for impurities scattering \cite{AGD}. This method can be modified \cite{Fal} 
in order to take  the possible localized states  into account.
Performing the iterations of Eq. (\ref{gfe}) in $U({\bf r})$
and averaging over the defect positions, we have 
 for the sum of noncrossing diagrams:
\begin{eqnarray} \label{sum}
D({\bf k},\omega)=D_0( k,\omega)+cD_0^2( k,\omega)\\ \nonumber
\times\sum_{n=0}^{\infty}(-u_0)^{n+1}\left(cD_0( k,\omega)+\sum_q
D( {\bf q},\omega)\right)^n,
\end{eqnarray}
which gives after some algebra  
\begin{equation} \label{def}
D({\bf k},\omega)^{-1}=D_0( k,\omega)^{-1}+cu_0
\left(1+u_0\sum_q D({\bf q},\omega)\right)^{-1}.
\end{equation}

The term of  first order only in the defect concentration $c$ 
is held in the phonon self-energy given by the second term on
the right-hand side of Eq. (\ref{def}), since we assume that
the distance between defects ($r_c\propto c^{-1/2}$ for line defects) is
larger, than the phonon mean-free path $r_{\gamma}$.

Let us emphasize that the phonon self-energy   involves the perturbed function
$D({\bf q},\omega)$, but not $D_0({\bf q},\omega)$. Then
the Green's function $D({\bf q},\omega)$ has to be found from
the integral equation (\ref{def}). Integration (summation) is perfomed
with respect to the 3D-vector ${\bf q}$ for point defects, over
the 2D-vector ${\bf q}_{\perp}$ for line defects (then $q_{z}=k_{z}$), 
and over
$q_{z}$ for plain defects (then ${\bf q}_{\parallel}={\bf k}_{\parallel}$).

The integral equation (\ref{def}) can be reduced to an algebraic
equation.
We introduce the new unknown function $\zeta$ and the complex
variable $\zeta_0$ 
in the case of point defects by the definition
$$D({\bf k},\omega)=(\zeta -s^2k^2)^{-1},\quad
\zeta_0=\omega_0^2-\omega^2-i\omega\Gamma^{\text{nat}}.$$
Then Eq. (\ref{def}) gives
\begin{equation} \label{sol1}
\zeta=\zeta_0+c\omega_0^3
\left(\frac{\kappa}{\lambda}-\kappa +\frac{\pi}{2}\sqrt{-\zeta}\right)^{-1}.
\end{equation}
 In the case of  line defects
$$D({\bf k},\omega)=(\zeta -s^2k_{\perp}^2)^{-1},\quad
\zeta_0=\omega_0^2-s^2k_{z}^2-\omega^2-i\omega\Gamma^{\text{nat}}$$ and 
we obtain
\begin{equation} \label{sol2}
\zeta=\zeta_0+c\omega_0^2\left(\frac{1}{\lambda}-
\ln{\frac{\kappa^2}{-\zeta}}\right)^{-1}.
\end{equation}
For  plain defects
$$D({\bf k},\omega)=(\zeta -s^2k_{z}^2)^{-1},\quad
\zeta_0=\omega_0^2-s^2k_{\parallel}^2-\omega^2-i\omega\Gamma^{\text{nat}},$$ 
we have
\begin{equation} \label{sol3}
\zeta=\zeta_0+c\omega_0^2\left(\frac{1}{\lambda}-
\frac{\omega_0}{\sqrt{-\zeta}}\right)^{-1},
\end{equation}
where $\lambda$ is the dimensionless coupling constant proportional to $u_0$,
$c$ is the atomic concentration  of defects in  3D, 2D, or 1D region,   
respectively for the point, line, or plain defects, 
$\kappa$ is the cutoff frequency, the $z$ axis  
 is taken in  direction of the line defect or normal
 to the plane defect, ${\bf k}_{\parallel}$ is parallel
 to the plane defect.
The branches of $\sqrt{-\zeta}$ and $\ln {(-\zeta)}$ have to be taken
in the upper half-plane. 
We consider the retarded Green's function which has no singularities
in the upper half-plane of  complex variable $\omega$. 

The equations (\ref{sol1})--(\ref{sol3}) have two singularities. 
One at $\zeta=0$
is related to the branch edge of the extended  optical-phonon mode.
 Other at $\zeta_l$ is
related to the localized mode and determined by the zero of 
expressions in  parentheses of Eqs. (\ref{sol1})--(\ref{sol3}).
For instance, for line defects we have
$$\frac{1}{\lambda}-\ln{\frac{\kappa^2}{-\zeta_l}}=0 $$
which gives
$$ \zeta_l=-\kappa^2e^{-1/\lambda}.$$
Notice, that the zero $\zeta_l$ of the expressions in parentheses,
if it exists,
is negative.  The zero can exist,
if $\lambda>1$ in the case of point  defects, and for any
positive $\lambda$ in the cases of both line  and plain defects.

The situation is schematically explained in Fig. 1, where
$\zeta_0$ determined by Eqs. (\ref{sol1})--(\ref{sol3}) is shown
 as  function of $\zeta$. In the regions 
  plotted by  solid lines  and indicated by the arrows 1,
 the solution $\zeta$  is  real  for  real values of
$\zeta_0$  and has no singularities in the
upper half-plain of $\omega$, as it must be for the
retarded Green's function. If $\Gamma^{\text{nat}}=0$, this region
makes no contribution to the density of phonon states. 
The second solution $\zeta$ (dashed lines indicated by the arrows 2) 
corresponding  to the same $\zeta_0$ should be rejected since it has
singularities  in the upper half-plain. 
For real $\zeta_0$,  the solution  $\zeta$ has the imaginary part 
in the regions shown by dot-dashed lines and indicated by arrows 3 and 4.
The region 3 represents the extended phonon mode. The region 4 shows the
localized states.

 a)  {\it  Low concentration of defects: unrenormalized approximation.} 

 We can solve  Eqs. (\ref{sol1})--(\ref{sol3})  analytically
 supposing $c$ to be small.
Using the iterative method (if $|\zeta-\zeta_0|\ll|\zeta_0|$)
we obtain on the first step  for point defects 
\begin{equation} \label{per1}
\zeta=\zeta_0+c\omega_0^3\left(\frac{\kappa}{\lambda}-\kappa +
\frac{\pi}{2}\sqrt{-\zeta_0}\right)^{-1},
\end{equation} 
 for line defects
\begin{equation} \label{per2} 
\zeta=\zeta_0+c\omega_0^2\left(\frac{1}{\lambda}-
\ln{\frac{\kappa^2}{-\zeta_0}}\right)^{-1},
\end{equation}
and for plain defects
\begin{equation} \label{per3}  
\zeta=\zeta_0+c\omega_0^2\left(\frac{1}{\lambda}-
\frac{\omega_0}{\sqrt{-\zeta_0}}\right)^{-1}.
\end{equation}
This unrenormalized approach corresponds to substituting
$D_0({\bf q},\omega)$ for $D({\bf q},\omega)$
in the self-energy of  Eq. (\ref{def}).

In  Eqs. (\ref{per1})--(\ref{per3}), the terms proportional to the 
concentration $c$ represent the contribution of defects. If  Re~$\zeta_0>0$, 
 i.e., for the region shown by the arrow 3 in Fig. 1, 
the frequency belongs 
to the extended states. Then
the imaginary and real parts of the second terms on the right-hand side of
Eqs. (\ref{per1})--(\ref{per3}) 
give the phonon shift and the additional width induced by
the defect - phonon scattering. 
If  Re~$\zeta_0<0$  and for positive
$\lambda$, the localized states can exist. In the absence of 
the natural width, they always exist for the cases both  the line  and plain
defects, in accord with the well-known statement of quantum mechanics 
about the bound state in a shallow potential.
However, for the finite value of $\Gamma^{\text{nat}}$, 
the coupling constant should be
strong  to give the substantial gap: $\omega_l-\omega_0\gg\Gamma^{\text{nat}}$.
This condition can be rewritten as $r_{ l} \ll r_{\gamma}$,
  if  we introduce the radius of 
 localized states $r_l=a\sqrt{\omega_0/(\omega_l-\omega_0)}$ and the
 phonon mean-free path $r_{\gamma}=a\sqrt{\omega_0/\Gamma^{\text{nat}}}$.


 b)  {\it Self-consistent approach: the square-root behavior.} 

The  conditions  of validity for
Eqs. (\ref{per1})--(\ref{per3}) impose a constraint
on the concentration of defects $c$.
 It is hard to satisfy
 the  conditions  of validity, 
 if $\zeta_0\rightarrow\zeta_l$. In fact, the equation
$\zeta_0=\zeta_l$ determines the frequency $\omega_l$
of localized phonon states.
For $|\zeta_0-\zeta_l|\ll |\zeta_l|$, we  expand the expressions in
 parentheses of Eqs. (\ref{sol1})--(\ref{sol3}) in ($\zeta-\zeta_l$)
and arrive to the quadratic algebraic equation 
which has the solution
\begin{equation} \label{qve}
\zeta=\frac{\zeta_0+\zeta_l}{2}+\sqrt{\frac{(\zeta_0-\zeta_l)^2}{4}-bc},
\end{equation}
where $b=4(-\zeta_l)^{1/2}\omega_0^3/\pi$ for point defects,
$b=-\zeta_l\omega_0^2$ for line defects, and 
$b=2\omega_0(-\zeta_l)^{3/2}$ for plain defects. 
We notice again that the solution
with the positive imaginary part should be taken in Eq. (\ref{qve}). 

If $\zeta_0\rightarrow\zeta_l$, we obtain a  negative value 
under the square root in 
Eq. (\ref{qve}), that gives the imaginary part of $\zeta$. 
 Then the imaginary part  is determined in  competition of
$\omega_0\Gamma^{\text{nat}}$ with $2(bc)^{1/2}$.
The region where $\Gamma^{\text{nat}}<2(bc)^{1/2}/\omega_0$ 
is indicated by the arrow 4 in Fig. 1. This condition may be compatible
with the conditions (i) $r_c\gg r_{\gamma}$ for the expansion in terms of $c$
 used in Eq. (\ref{def}) and (ii) $r_{\gamma}\gg r_l$ for 
 the existence of localized states.  
 Here,  the localized states make the maximal contribution 
to the differential density of states and to the Raman intensity,
\begin{equation} \label{wid} 
  \text{Im}~\frac{1}{\zeta}=\frac{\sqrt{bc}}{\zeta_l^2}, 
\end{equation}
which is proportional to the square root of the defect concentration.
 In the opposite limit 
$\Gamma^{\text{nat}}>2(bc)^{1/2}/\omega_0$, we can use the Eqs. 
(\ref{per1})--(\ref{per3}) which give the contribution 
linear in the defect concentration.

 c) {\it Self-consistent approach: the numerical solution.}

 As a matter of fact, each of the equations (\ref{sol1}), (\ref{sol2}), and 
 (\ref{sol3}) is a system of  two equations for the real and imaginary
 parts of $\zeta$. This system may be solved numerically.
 It is convenient to introduce the  unknown real variables $x$ and $y$ as 
 $$-\zeta= \sqrt{x^2+y^2}\exp 
 {\left[i\left(\frac{\pi}{2}-\arctan\frac{x}{y}\right)\right]}$$
 supposing $y>0$ and $-\pi/2<\arctan z<\pi/2$ in order to comply with the   
 analytical requirements in the upper-half plain. The similar form is used
 also for $ \zeta_0$ with $x\rightarrow\omega^2-\omega_0^2$ and 
 $y\rightarrow\omega\Gamma^{\text{nat}}$.
 
 The results of the numerical solution are shown by solid lines in
 Fig. 2  (line defects) and  in
 Fig. 3  (plain defects) for the cases, when the coupling constant $\lambda$
 is negative and no localized states exist. We used the  values
 of  $\omega_0=520$ cm$^{-1}$ and $\Gamma^{\text{nat}}=3.2$ cm$^{-1}$, which
 correspond to pure silicon, and the cutoff parameter $\kappa=\omega_0$.
  The  unrenormalized approximation is shown by dashed lines. 
 We see that the interaction of phonons
 with defects results in  broadening and  shift of the Raman line 
 from the position (dotted line) in the absence of defects.
   
 The broadening is much more on the low-frequency side of the line
 than on the high-frequency side, that gives the line shape asymmetry.
 This is an effect of the phonon density of states: the elastic interaction
 with defects makes a contribution to the phonon life-time only in the
 region $\omega^2<\omega_0^2$, where the final phonon states exist.
 
 In Fig. 4, the dependences of the line shift and width on the concentration
 are shown for the point, line, and plain defects. The value of 
 coupling constant $\lambda=-1$ is taken. We define 
 the full width at half maximum $\Gamma$ (FWHM) in the 
 usual way: $ \Gamma=\omega_{+}-\omega_{-}$, where the Raman intensities 
 $I(\omega_{\pm})=
 0.5I(\omega_{\text{max}})$ and $\omega_{+}>\omega_{\text{max}}>\omega_{-}$.
 The behavior of the width $ \Gamma$ and  of the line position
 $\omega_{\text{max}}$, calculated with the help of Eqs. (\ref{sol1}), 
 (\ref{sol2}), and 
 (\ref{sol3}) is shown in the left panel. In the right panel, the Raman line
 asymmetry $(\omega_{+}-\omega_{\text{max}})-
 (\omega_{\text{max}}-\omega_{-})$ is plotted. We see that the plain defect
 affect more intensively on the width and asymmetry,
   than the line and point defects. To the contrary, 
  the Raman line is shifted more for the point defects.
 
 The cases of the positive coupling constant, when the localized states
 arise, are shown in Figs. 5 and 6. The accurate results (solid lines)
 differ considerably from that obtained in the unrenormalized approximation
 (dashed lines). The maximum that was found in the unrenormalized approximation
 is smoothed out, transforming into a shoulder with the width of the order of
 $\sqrt{c}$ in accord with Eq. (\ref{qve}). Emphasize, that the total
 number of localized states, integrated over frequency, is proportional to
 $c$ independently of the approximation. 
 
 The difference in results for
 defects of different geometries is evident from Figs. 2--6. 
 These distinctions in the Raman line shape
and shift can be detected in fitting to the experimental data.
 \section {Conclusions} 
To summarize, we investigate here the structure of the phonon density
of states and the Raman line shape 
in the situation when  localized levels $\omega_l$
 interact strongly with the closely-spaced edge of the extended 
 optical-phonon states.
 In the linear approximation in defect concentration $c$, 
 the contribution of localized states does not have the well-known form
of $\delta$-function: $c \delta (\omega-\omega_l)$, if the effect of 
defect concentration is more essential  than the phonon natural width. 

The line shape of bulk-phonon extended states is asymmetrical.  
The additional width (proportional to the defect concentration)
arises on the side of line related to the band of extended states. As a
most intriguing example, we would like to compare the behavior of
the Raman lines   of graphite at 1600 cm$^{-1}$ and
 of diamond-like material at 1300 cm$^{-1}$. 
 When defects are added to graphite, the
 low-frequency wing of the line becomes to drop more slowly than the
 high-frequency wing (see Figs. 1 and 2 in Ref. \cite{WRW}) in just the
 same way as shown in Figs. 2 and 3 of the present paper. But if it is diamond
 samples that are impaired, the opposite, i.e. high-frequency,  side slopes
 more gently (see Fig. 7  in Ref. \cite{WRW} and  Fig. 2  in Ref. \cite{AWR}).
 Using the present theory, the explanation of the phenomenon  may be
 attributed to the difference in the phonon dispersion of these substances.
 Namely, contrary to graphite, there is a minimum of the optical-phonon branch 
 at the Brillouin zone-center in diamond \cite{SMA}. 
\section{acknowledgments}
 
The author acknowledges the kind hospitality of Max-Plank-Institute
f\"{u}r Physik Komplexer Systeme, Dresden, 
 where this work was performed.

\newpage
\begin{figure}[tb]
\label{1}
 \epsfxsize=120mm
 \epsfysize=120mm
\centerline{\epsfbox{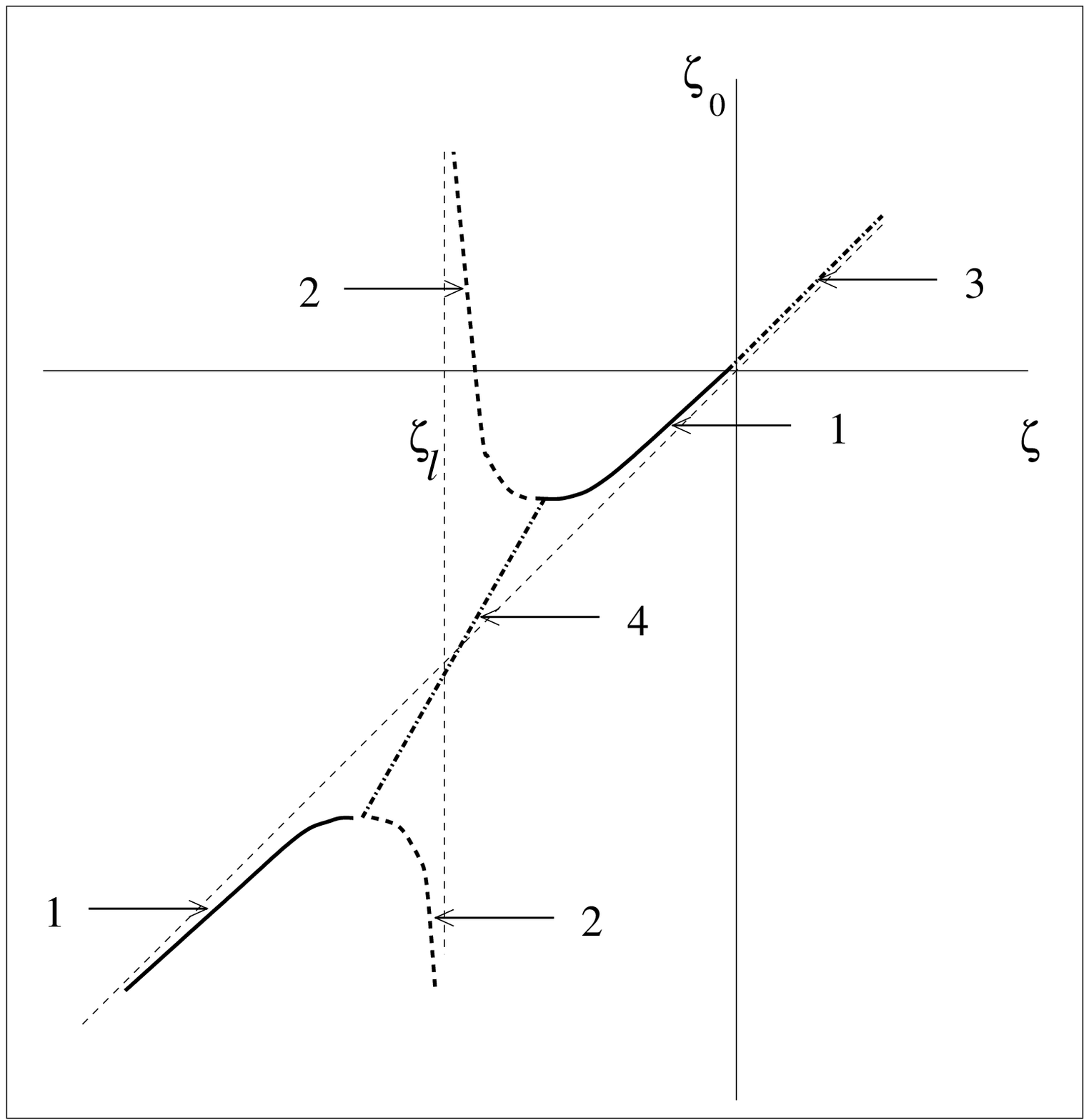}}
\caption[]{ 
Plot of function $\zeta_0(\zeta)$  given by Eqs. (\ref{sol1})--(\ref{sol3}).
 In the regions shown by the dot-dashed  lines (indicated by arrows 3 and 4), 
the solution  
$\zeta(\zeta_0)$ has a finite imaginary parts for  real $\zeta_0$.
The region 3 corresponds to the extended  phonon mode. The region
4 is related to the localized states.}
\end{figure}

 \newpage
 \begin{figure}[tb]
\label{2}
 \epsfxsize=100mm
 \epsfysize=140mm
\centerline{\epsfbox{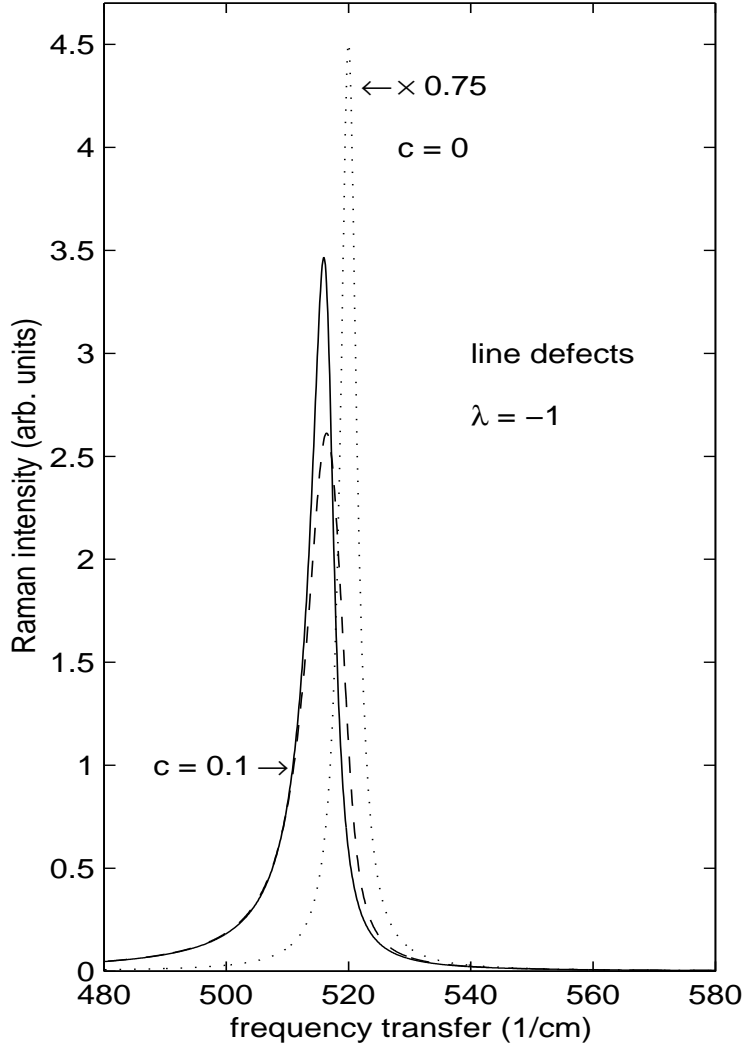}}
\caption[]{Imaginary part of the phonon Green's function versus
 frequency $\omega$ for  parameters of silicon: $\omega_0=520$ cm$^{-1}$,
  $\Gamma^{\text{nat}}=3.2$ cm$^{-1}$. The number of line defect  
  per atomic volume $c=0.1$ and the phonon-defect
  coupling constant $\lambda = -1$ 
  (the solid line for the self-consistent solution, 
  the dashed line for unrenormalized approximation). The results for pure
  crystal ($c=0$) are shown by the dotted line.}
\end{figure}

\newpage
 \begin{figure}[tb]
\label{3}
 \epsfxsize=100mm
 \epsfysize=140mm
\centerline{\epsfbox{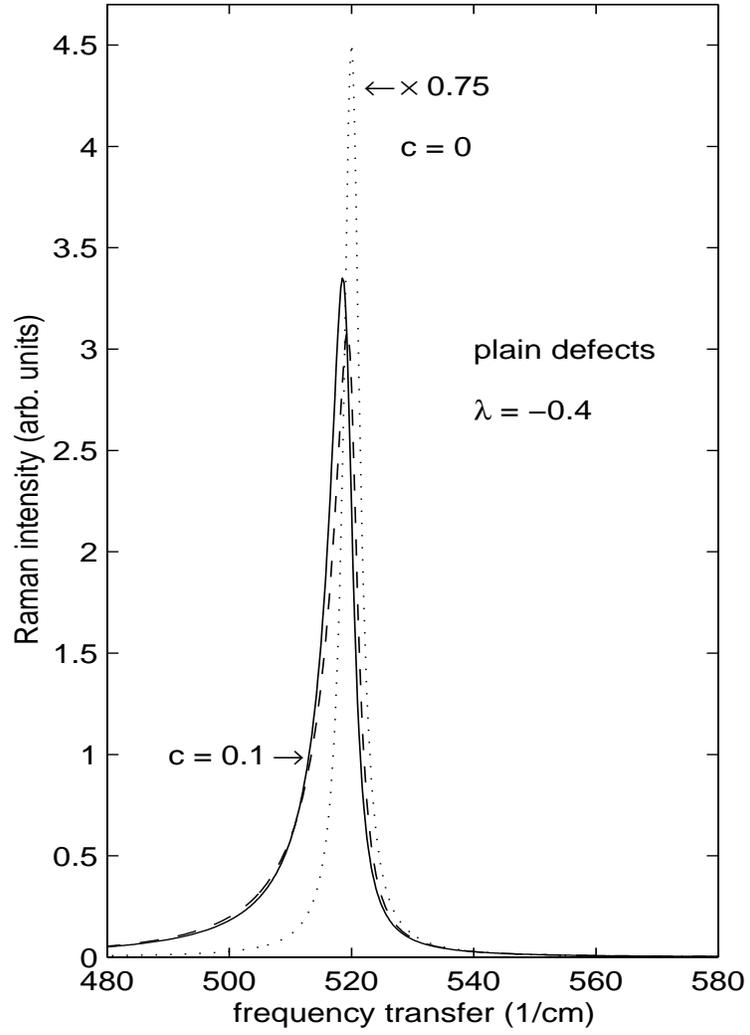}}
\caption[]{
Same as Fig. 2 but for plain defects with the coupling constant
 $\lambda = -0.4$.} 
\end{figure}

\newpage
 \begin{figure}[tb]
\label{4}
 \epsfxsize=140mm
 \epsfysize=120mm
\centerline{\epsfbox{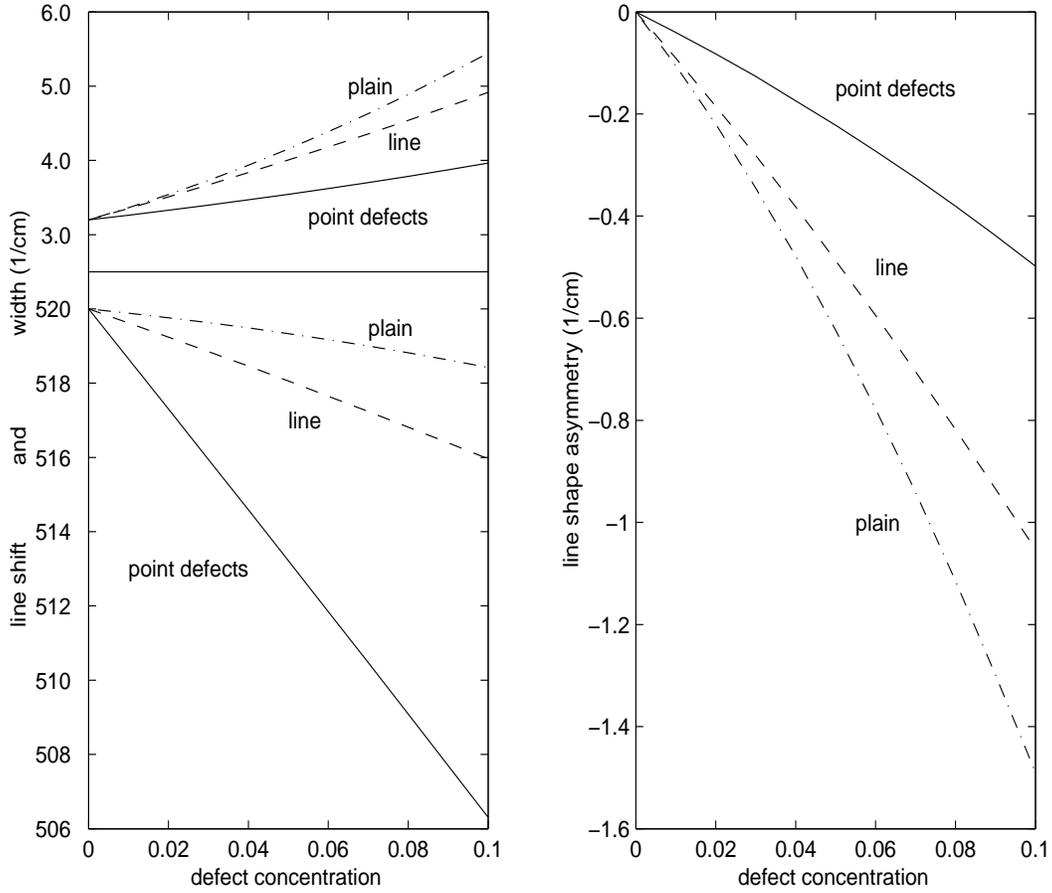}}
\caption[]{
The theoretical Raman line width (upper part of left panel), shift 
(bottom part of left panel), and asymmetry
(right panel) as functions of the defect concentration. 
The phonon-defect coupling constant
 $\lambda = -1$. Results for the point, line, and plain defects are
 plotted by the solid, dashed, and dot-dashed lines, respectively.} 
\end{figure}

\newpage
 \begin{figure}[tb]
\label{5}
 \epsfxsize=100mm
 \epsfysize=140mm
\centerline{\epsfbox{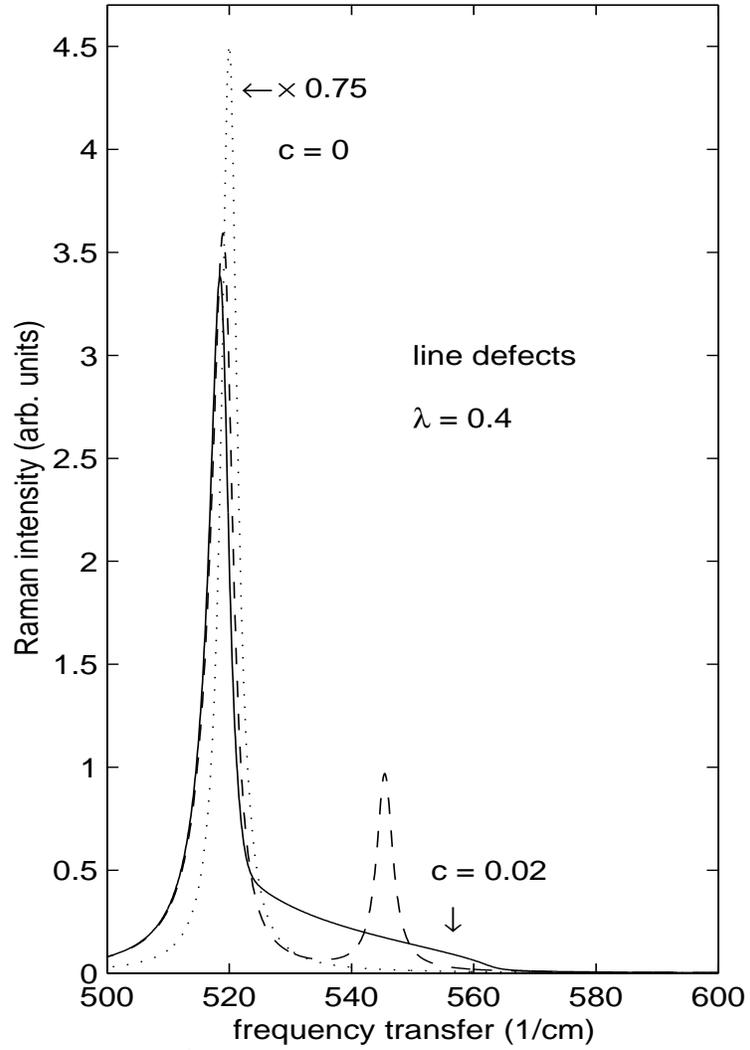}}
\caption[]{
Same as Fig. 2 but for the positive coupling constant
 $\lambda = 0.4$, when localized states exist. The defect concentration 
  $c=0.02$.} 
\end{figure}

\newpage
 \begin{figure}[tb]
\label{6}
 \epsfxsize=100mm
 \epsfysize=140mm

\centerline{\epsfbox{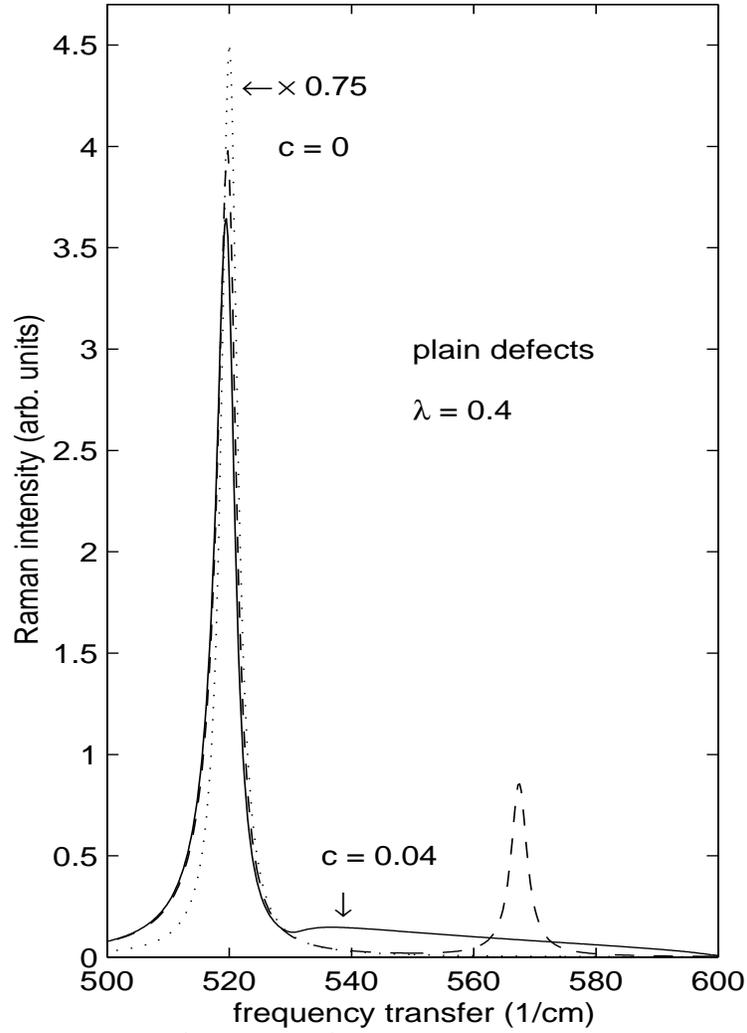}}
\caption[]{
Same as Fig. 2 but for plain defects with the positive coupling constant
 $\lambda = 0.4$ and the  concentration  $c=0.04$.} 
\end{figure}
\end{document}